\documentclass[10pt,twocolumn,twoside,letterpaper]{IEEEtran}

\usepackage{geometry}
\usepackage{cite}
\usepackage{amsmath,amssymb,amsfonts}
\usepackage{graphicx}
\usepackage{siunitx}
\graphicspath{{images/}}

\makeatletter
\def\ps@IEEEtitlepagestyle{
  \def\@oddfoot{\mycopyrightnotice}
  \def\@evenfoot{}
}
\def\mycopyrightnotice{
  {\footnotesize
  \begin{minipage}{\textwidth}
  \centering
  978-1-6654-2113-3/21/\$31.00 \copyright2021 IEEE
  \end{minipage}
  }
}

\usepackage{url}

\begin{document}
\title{Rate capability of large-area triple-GEM detectors and new foil design for the innermost station, ME0, of the CMS endcap muon system}
%
%
%

\author{M. Bianco$^a$,
F. Fallavollita$^a$,
D. Fiorina$^b$,
A. Pellecchia$^{1,c}$,
L. F. Ramirez Garcia$^d$,
N. Rosi$^b$,
P. Verwilligen$^c$
on behalf of the CMS Muon Group 
\thanks{$^1$ Corresponding author}
\thanks{$^a$ CERN, Switzerland}
\thanks{$^b$ University and INFN Pavia, Italy}
\thanks{$^c$ University and INFN Bari, Italy}
\thanks{$^d$ Universitad de Antioquia, Columbia}
\thanks{Manuscript received December 1, 2021}
}

\maketitle

\pagenumbering{gobble}

\begin{abstract}
To extend the acceptance of the CMS muon spectrometer to the region 2.4$<|\eta|<$2.8, stacks of triple-GEM chambers, forming the ME0 station, are planned for the CMS Phase 2 Upgrade. These large-area micro-pattern gaseous detectors must operate in a challenging environment with expected background particle fluxes up to \textbf{\SI{150}{\kilo\Hz\per\centi\m^2}}. Unlike traditional non-resistive gaseous detectors, the rate capability of such triple-GEM detectors is limited not by space charge effects, but by voltage drops on the chamber electrodes due to avalanche-induced currents flowing through the resistive protection circuits (introduced as discharge quenchers). We present a study of the irradiation of large-area triple-GEM detectors with moderate fluxes to obtain a high integrated hit rate. The results show drops as high as 40\% of the nominal detector gas gain, which would result in severe loss of tracking efficiency. We discuss possible mitigation strategies leading to a new design for the GEM foils with electrode segmentation in the radial direction, instead of the “traditional” transverse segmentation. The advantages of the new design include uniform hit rate across different sectors, minimization of gain-loss without the need for voltage compensation, and independence of detector gain on background flux shape.
\end{abstract}

\section{Introduction}
\label{sec:introduction}
\begin{figure}[h]
    \centering
    \includegraphics[width=0.5\textwidth]{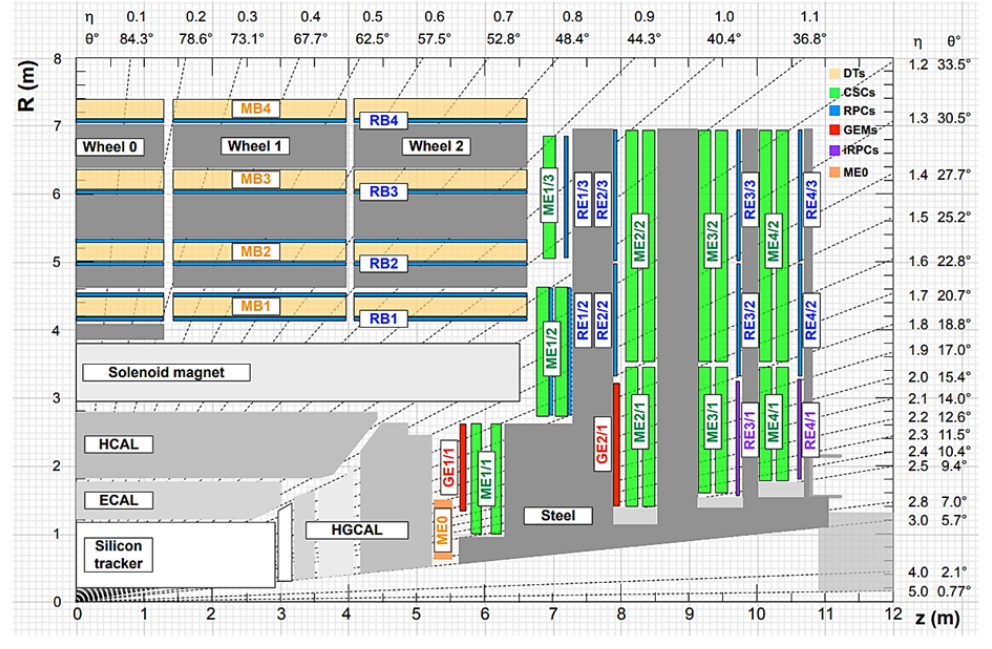}
    \caption{Quadrant of the CMS muon spectrometer after the planned upgrade for HL-LHC, including the ME0 station in orange.}
    \label{fig:me0}
\end{figure}

\begin{figure}[h]
    \centering
    \includegraphics[width=0.4\textwidth]{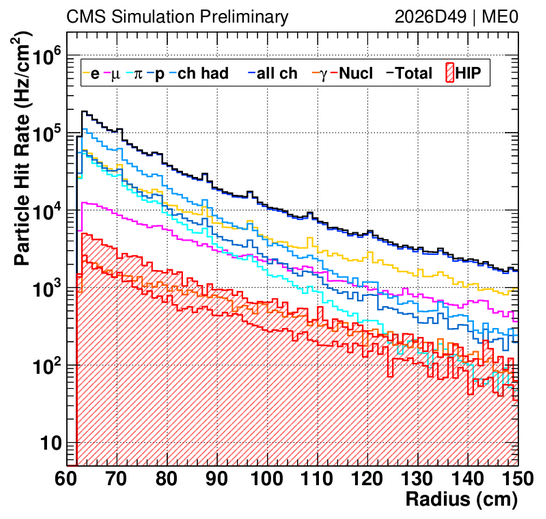}
    \caption{Expected background flux in the ME0 environment as a function of the distance from the LHC beam line.}
    \label{fig:bkg}
\end{figure}

\IEEEPARstart{T} ~he ME0 station of the CMS experiment\cite{cms} is planned to be installed for the CMS Muon Spectrometer upgrade (Fig.~\ref{fig:me0}) to cope with the increased pileup in the high-luminosity LHC upgrade and to extend the CMS muon system coverage to 2.4$<|\eta|<$2.8. The requirements for the detector technology chosen for the instrumentation of the ME0 station \cite{tdr} include space resolution under \SI{500}{\micro\radian}, time resolution under \SI{10}{\nano\s} and longevity over an integrated charge of \SI{9}{C\per\centi\m^2}. The background particle flux expected in the ME0 station in the CMS environment during the LHC operations, initially estimated to be lower than \SI{50}{\kilo\Hz\per\centi\m^2}, has been updated after a more precise redesign of the CMS detector geometry to a maximum value of \SI{150}{\kilo\Hz\per\centi\m^2} in the highest pseudorapidity region, with exponential drop at increasing distance from the LHC beam line (Fig. \ref{fig:bkg}).

The triple-GEM technology is well-suited for the ME0 station among all micro-pattern gaseous detectors with its good space and time resolutions (up to respectively ~150 µm and 7-8 ns) \cite{fabbiucc}. The rate capability of triple-GEM detectors has been verified with X-rays up to \SI{100}{\mega\Hz\per\milli\m^2} on irradiated areas of few \SI{}{\milli\m^2} \cite{pieter}. Such measurements are mainly sensitive to local effects such as the ion space charge effect, a distortion of the electric field in the detector gas volume due to the high density of slowly-moving positive charges in the gas \cite{patrick}.

\begin{figure}[h]
    \centering
    \includegraphics[width=0.4\textwidth]{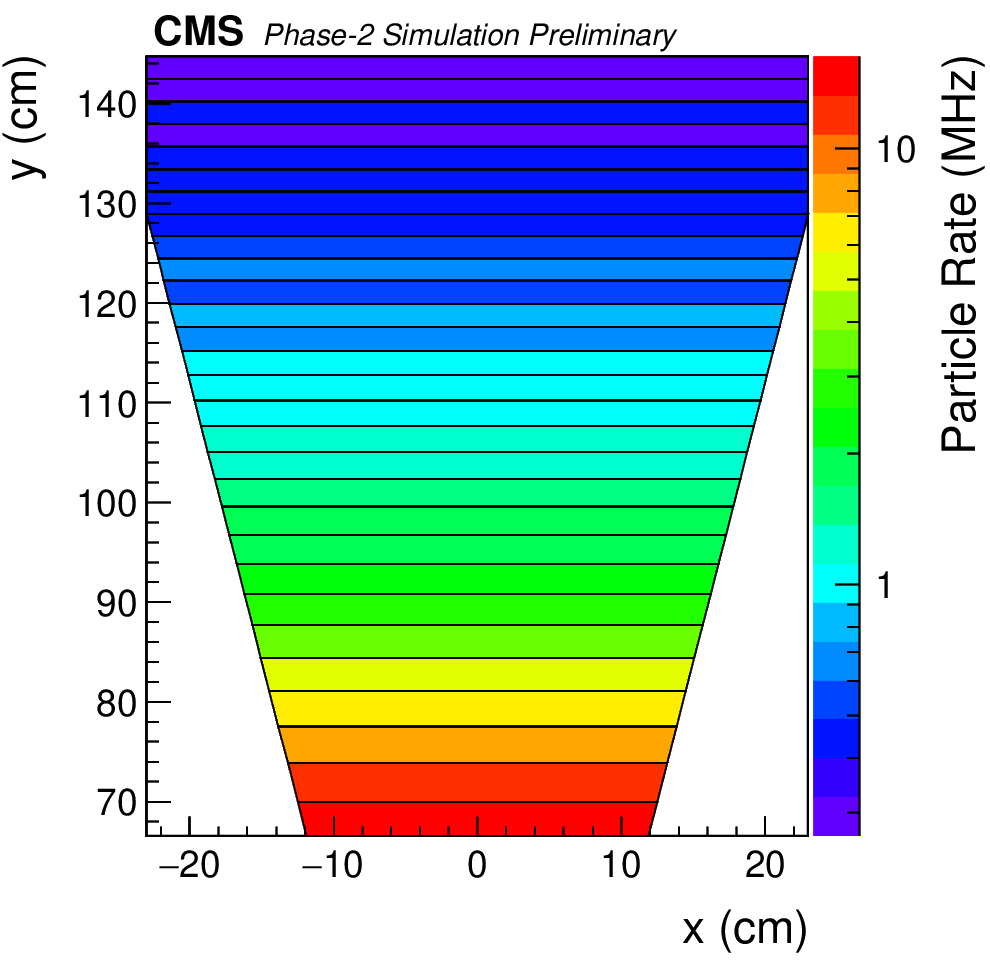}
    \caption{Scheme of an ME0 detector with transverse GEM foil segmentation with respect to the LHC beam line, showing the expected background particle rate per sector.}
    \label{fig:me0_transverse_segmentation}
\end{figure}

In their final design, the ME0 detectors are equipped with segmented GEM foils \cite{ciccio} (Fig.~\ref{fig:me0_transverse_segmentation}), each sector (of area \SI{100}{\centi\m^2}) being connected to the power supply through a quenching resistor to reduce the energy of possible discharges.

\begin{figure}[h]
    \centering
    \includegraphics[width=0.4\textwidth]{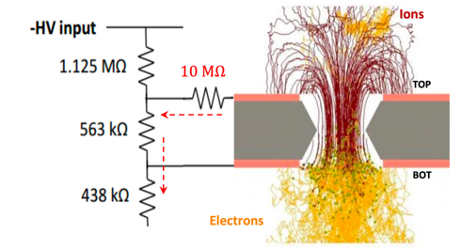}
    \caption{Scheme of the avalanche-induced currents flowing through the protection resistors of a single GEM foil.}
    \label{fig:avalanche}
\end{figure}

When the detector is irradiated, the charges produced in the avalanche and moving in the gas induce currents on the electrodes that flow through the protection circuits (Fig.~\ref{fig:avalanche}), resulting in a voltage drop on the electrode itself; the voltage drop is a global effect over the entire foil, i.e. it increases for wider irradiated areas. As a consequence, the rate capability of the CMS triple-GEM detectors should be re-evaluated in a measurement at high integral hit rates, obtained by irradiating the entire detector surface at moderate particle flux. The rate capability measurement is described in the following sections, followed by a comparison of its results with the ME0 requirements and a discussion of the chosen mitigation strategies.

\section{Rate capability measurement}
\label{sec:setup}
\begin{figure}[h]
    \centering
    \includegraphics[width=0.5\textwidth]{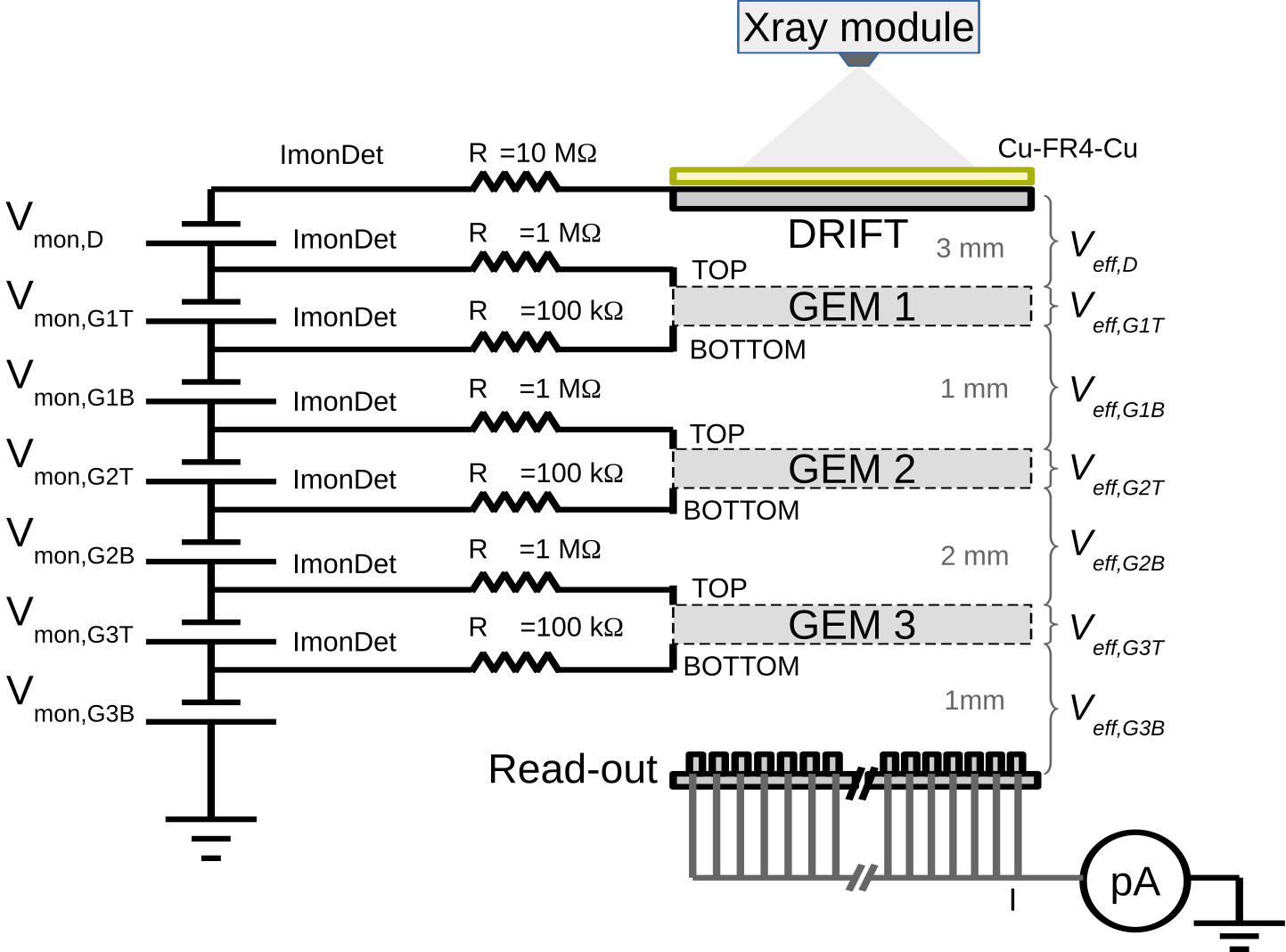}
    \caption{Rate capability measurement setup. The detector, equipped with protection resistors and irradiated with two x-ray sources, is powered through a multi-channel power supply and read out on the anode by a picoammeter.}
    \label{fig:rate_capability_setup}
\end{figure}

The rate capability measurement has been performed on a triple-GEM detector of 10x\SI{10}{\centi\m^2} area irradiated simultaneously by two silver-target x-ray generators \cite{amptek} at increasing fluxes (Fig.~\ref{fig:rate_capability_setup}). Within a single measurement, the distance of the source from the detector was gradually decreased to span a wider range of particle rates from 200 kHz to 20 MHz.

The detector under study was powered to a nominal gas gain of \SI{2e4}{} through a CAEN A1515 power supply, that allows monitoring of the currents on the cathode and GEM electrodes with an accuracy of \SI{100}{\pico\ampere}. The anode current was measured by an ammeter of resolution 1 pA.

\subsection{Rate measurement}

The typical rate of traditional laboratory counting electronics is limited by pile-up to about \SI{1}{\mega\Hz}, lower than the maximum particle rate achievable in the setup. Therefore, the hit rate was extrapolated from the anode current (measurable with good linearity up to \SI{20}{\milli\ampere}) with a procedure outlined below.

\begin{figure}[h]
    \centering
    \includegraphics[width=0.5\textwidth]{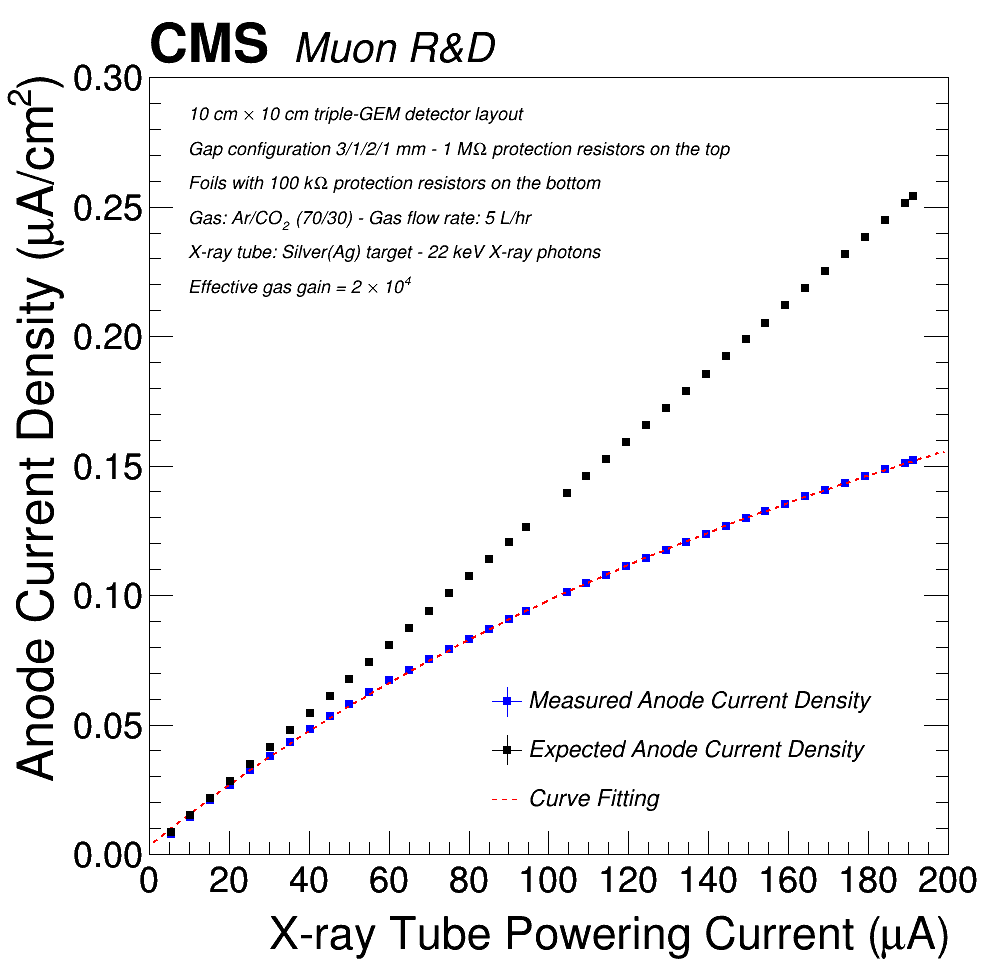}
    \caption{Anode current of the detector as a function of the x-ray tube powering current (blue) and linearized anode current (black).}
    \label{fig:anode_current_vs_xray}
\end{figure}

The anode current was measured with increasing flux of the x-ray sources (Fig.~\ref{fig:anode_current_vs_xray}). The saturating trend is due to the gain drop of the detector. The current curve was fitted with the parametric function
\begin{equation}
    i_\text{anode} = \frac{A P_\text{xray}+ B}{1+k(A P_\text{xray}+ B)},
    \label{eq:anode}
\end{equation}
(where $P_\text{xray}$ is the current powering the x-ray source and $k$ is a parameter quantifying the bending of the curve) and then linearized to obtain the anode current that would be measured on the detector in absence of gain drop.

\begin{figure}[h]
    \centering
    \includegraphics[width=0.5\textwidth]{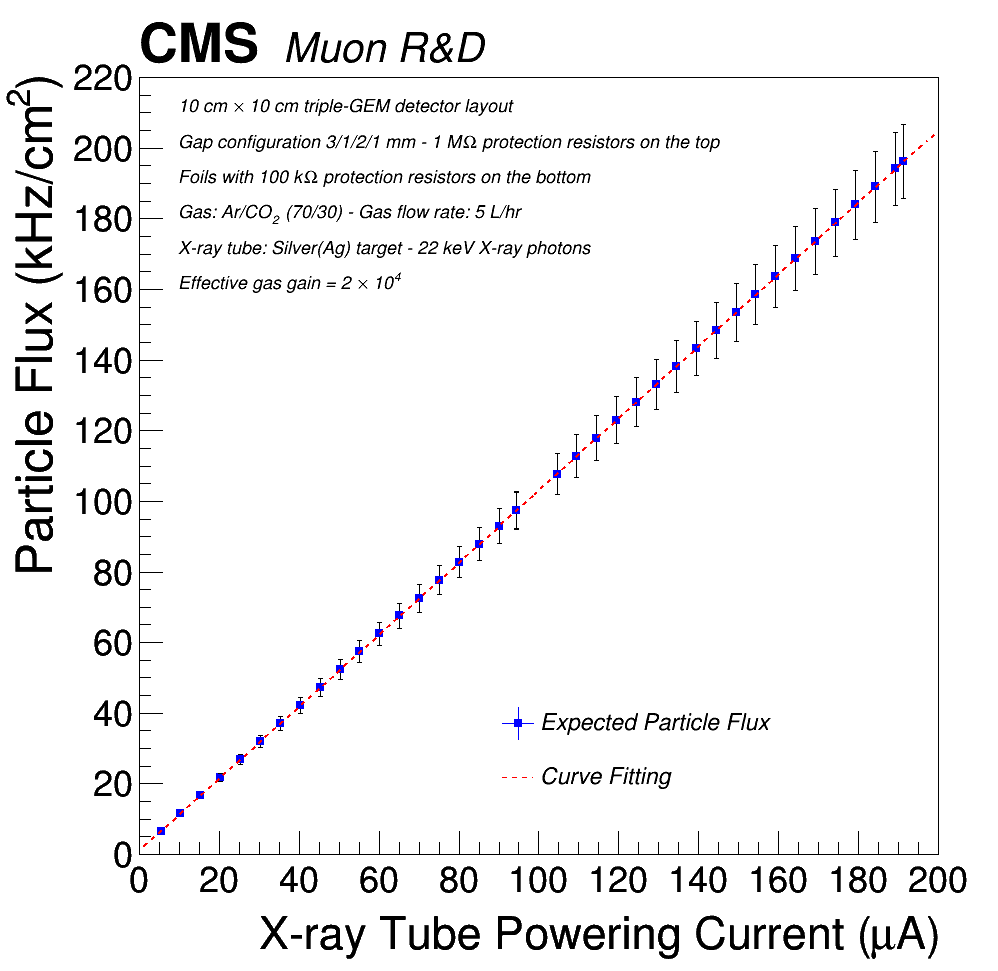}
    \caption{Hit rate on the detector as a function of the x-ray current obtained by inversion of the effective gain equation.}
    \label{fig:rate_vs_xray}
\end{figure}

The rate on the chamber (Fig.~\ref{fig:rate_vs_xray}) was finally calculated from the linearized current by inversion of the effective gain equation:
\begin{equation}
    R = \frac{i_\text{linearized}}{q_e n_p g_\text{nominal}},
    \label{eq:rate}
\end{equation}
where $q_e$ is the electron charge, $n_p$ is the average number of primary electrons created by the x-ray photons in the chamber per event and $g_\text{nominal}$ is the detector effective gain in absence of background flux.

\subsection{Gain measurement}

The effective gain of the detector under irradiation was determined separately -- for each rate point -- by two independent methods.

The first gain estimation (`direct' method) is obtained by the effective gain equation:
\begin{equation}
    g_\text{eff} = \frac{i_\text{anode}}{q_e n_p R},
    \label{eq:eff_gain_direct}
\end{equation}
where $R$ is the particle rate as determined with the procedure described above. The main source of uncertainty in this measurement is the rate extrapolation error.

In the second gain measurement, each $i$-th electrode is powered at an `effective' voltage obtained from the nominal voltage supplied by the power module by subtracting the voltage drop on the $i$-th protection resistor:
\begin{equation}
    V^i_\text{eff} = V^i_\text{set}-I^i_\text{mon}R^i_\text{protection}.
    \label{eq:veff}
\end{equation}

A gain measurement is performed at such "effective" bias point at low irradiation by an x-ray source, measuring the particle rate with a preamplifier and a scaler. The gain estimated with this procedure is more accurate than the former because it is a direct measurement, but it is only sensitive to gain variations traceable to the `ohmic' effect, i.e. the voltage drop on the protection resistors.

\subsection{Results}

\begin{figure}[t]
    \centering
    \includegraphics[width=0.5\textwidth]{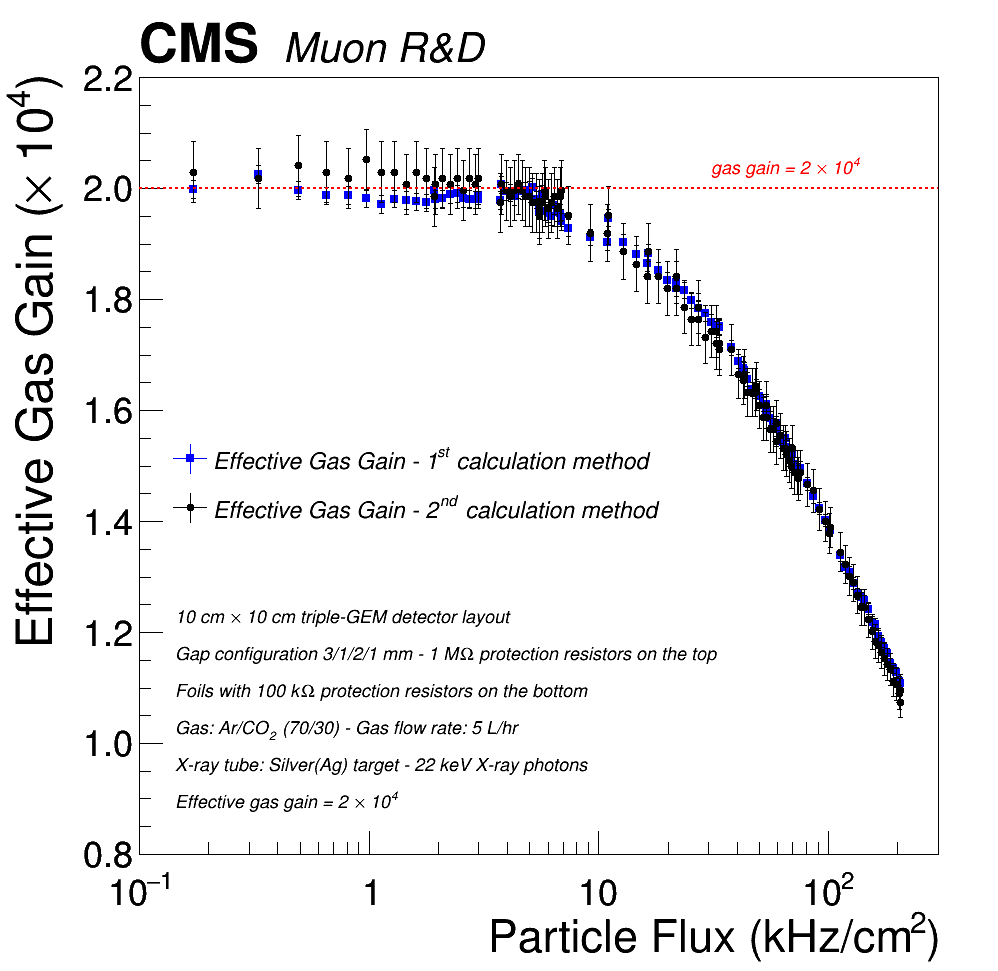}
    \caption{Gain drop of the irradiated prototype comparing the `direct' gain measurement (in black) and the `ohmic' gain measurement (in blue).}
    \label{fig:rate_capability_10x10}
\end{figure}

The rate capability curves (Fig.~\ref{fig:rate_capability_10x10}) obtained with the `direct' and `ohmic' experimental methods are in agreement within the measurement statistical uncertainty, proving that the only perceptible gain drop in the explored rate interval is due to the already discussed `resistive' effect.

The highest gain drop observed in the laboratory measurements is 40\% of the expected gas gain, observed at an irradiation of \SI{100}{\kilo\Hz\per\centi\m^2}. However, such value depends on the irradiated detector surface and the values of the protection resistors on each electrode.

\section{Gain recovery by voltage compensation}

A compensation measurement was performed to determine the new bias voltage at which the detector should be powered, for a set irradiation rate, to recover the original nominal gain of $2\times 10^4$ while maintaining a fixed ratio between the  gap fields and the GEM foil voltage differences. The measurement was carried out iteratively by increasing the applied voltage on each GEM electrode until the effective voltage on each electrode (calculated from Eq.~\ref{eq:veff}) was equal to the voltage required to operate the detector at the nominal gain.

\begin{figure}[t]
    \centering
    \includegraphics[width=0.45\textwidth]{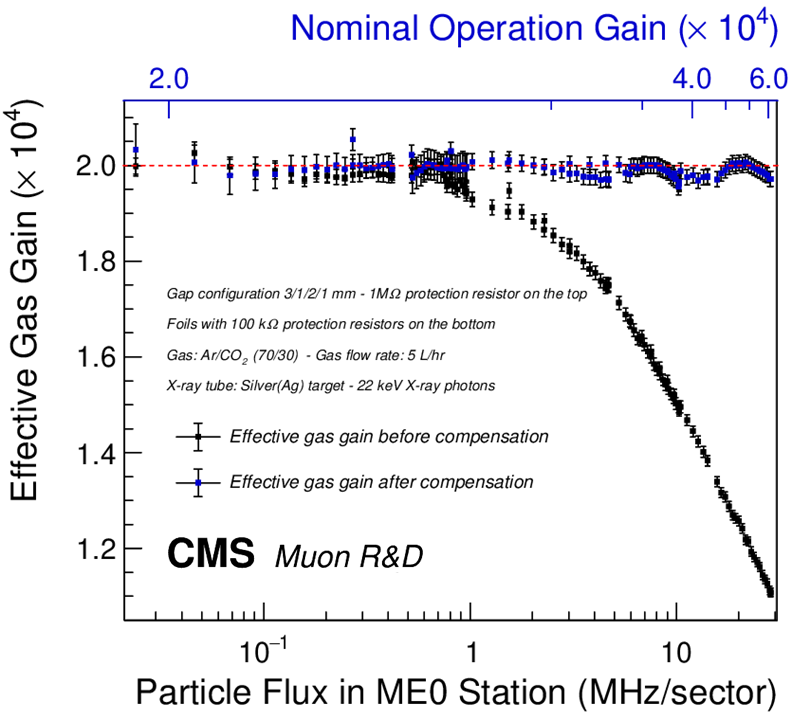}
    \caption{Recovered gain of the GEM detector over the tested rate region. The top axis shows the nominal gain (as per gain calibration curve) at which the detector has to be powered to operate it at an effective gain of \SI{2e4}{} under irradiation.}
    \label{fig:benciplot}
\end{figure}

Fig.~\ref{fig:benciplot} shows the rate capability curve overlapping with the recovered gain curve after voltage compensation, together with the nominal gain at which the detector has to be powered to operate it at an effective gain of \SI{2e4}{} for each value of the background rate. The gain is expressed as a function of the integral hit rate on the detector (instead of the particle flux) to make the result independent of the irradiated area. The rate axis has also been rescaled to keep into account the different average number of primary electrons per event expected in the ME0 environment in CMS compared to the x-ray source.

Despite proving that a gain compensation by overvoltage is achievable, the result shows that an effective operation of the ME0 detector in the highest-$\eta$ region (see Fig.~\ref{fig:me0_transverse_segmentation}) would require a nominal gain of approximately \SI{6e4}{}, with an increased damage risk in case of sudden beam loss during the LHC operation. Moreover, each GEM foil sector would have to be powered at a different voltage point to account for the highly uneven background rate at increasing pseudorapidity. The proposed solution for the two issues raised above is discussed in the next section.


\section{Azimuthal segmentation of the GEM foils}

\begin{figure}[t]
    \centering
    \includegraphics[width=0.45\textwidth]{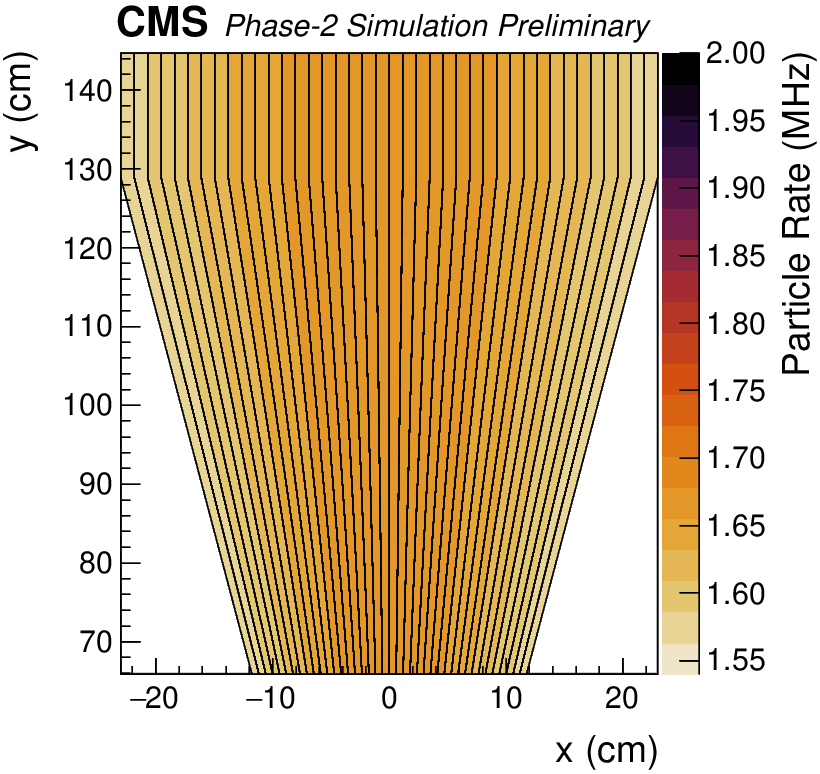}
    \caption{Design of the adopted azimuthal segmentation for the ME0 detectors, showing the expected background particle rate per sector in the CMS environment.}
    \label{fig:phi_segmentation}
\end{figure}

\begin{figure}[h]
    \centering
    \includegraphics[width=0.4\textwidth]{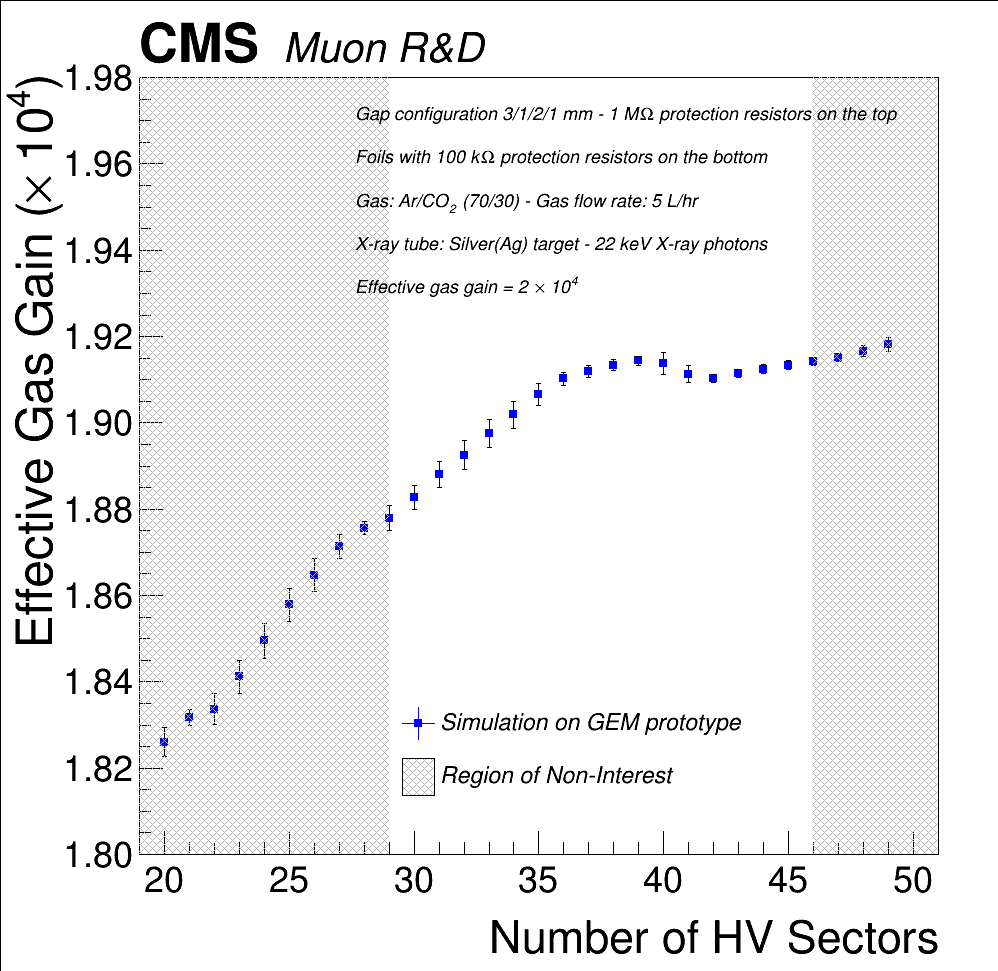}
    \caption{Expected average gain drop of the ME0 detectors for different number of azimuthal foil segments.}
    \label{fig:gain_vs_segmentation}
\end{figure}

\begin{figure}[h]
    \centering
    \includegraphics[width=0.45\textwidth]{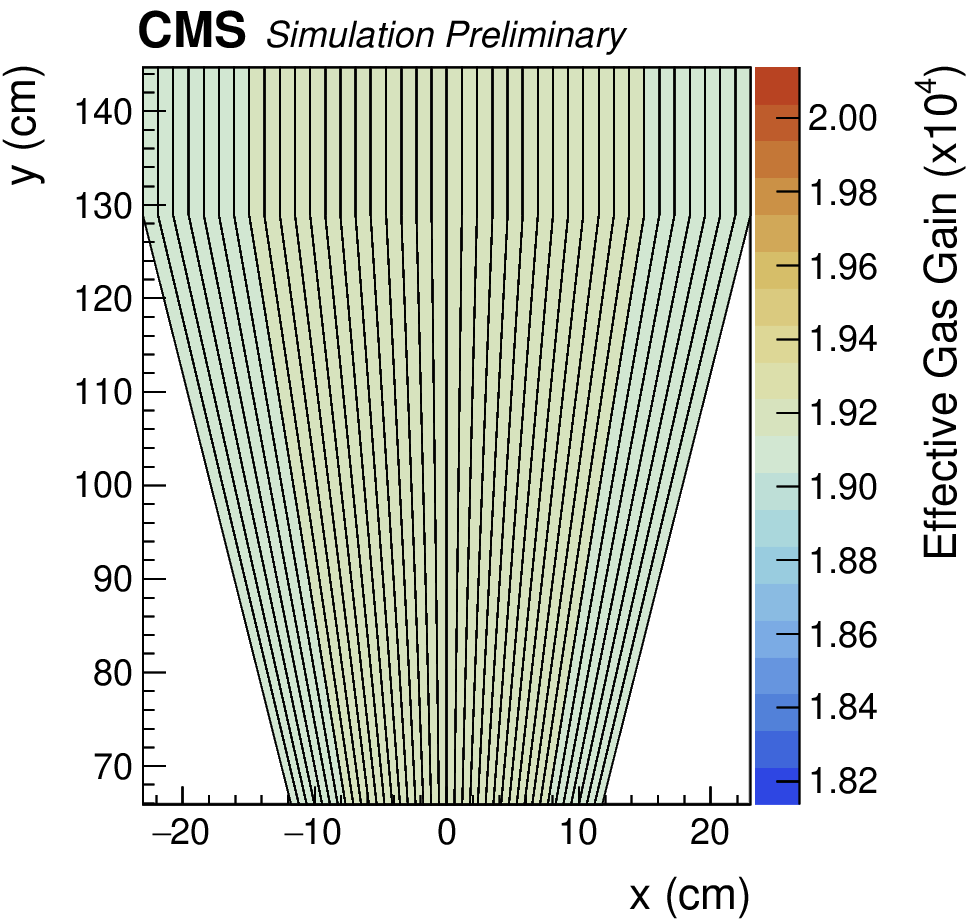}
    \caption{Expected gain of the ME0 detectors with azimuthal segmentation for each GEM foil sector under irradiation in the CMS environment.}
    \label{fig:phi_gain}
\end{figure}

A possible design choice for equalizing the gain drop across different GEM foil sectors and minimizing the average gain drop consists of abandoning the longitudinal foil segmentation in favor of an azimuthal sectorization wth respect to the beam axis (Fig.~\ref{fig:phi_segmentation}). By combining the background flux simulations and the gain drop measured in the rate capability studies it is possible to determine an expected dependency of the gain drop of the ME0 detector with azimuthal segmentation as a function of the number of sectors (Fig.~\ref{fig:gain_vs_segmentation}). In a segmentation with 40 sectors (Figs.~\ref{fig:phi_segmentation} and \ref{fig:phi_gain}) the average rate per HV sector in the CMS background can be contained to 1.5 MHz, while the effective gain drop can be minimized to about 10\% of the expected value of \SI{2e4}{}.

An essential aspect to be considered in the comparison between the longitudinal and the azimuthal segmentation for future studies involves the different dead area introduced by the polyimmide separation lines between two adjacent foil sectors. Considering that each separation line has a thickness of \SI{200}{\micro\m}, the total dead area of a ME0 module with 40-sector azimuthal segmentation is approximately 1.75\% of the detector surface, compared with about 0.1\% for a 29-sector longitudinally segmented detector.

\section{Conclusions}
A new approach to the rate capability problem of triple-GEM detectors has been applied for the high-rate environment expected for the innermost muon station of the CMS endcaps for the high-luminosity upgrade. The rate capability of large-area triple-GEM detectors has been shown to be limited by the protection circuits applied to the detectors and to be independent of the irradiated area at fixed hit rate. The measured gain drops can be as high as 40\% of the expected gain, which can be recovered by applying overvoltages to the detector electrodes. A mitigation strategy chosen for the CMS detectors involves an azimuthal segmentation of the ME0 chambers, expected to limit the gain loss during CMS operations to less than 10\%.

\begin{figure}[h]
    \centering
    \includegraphics[width=0.35\textwidth]{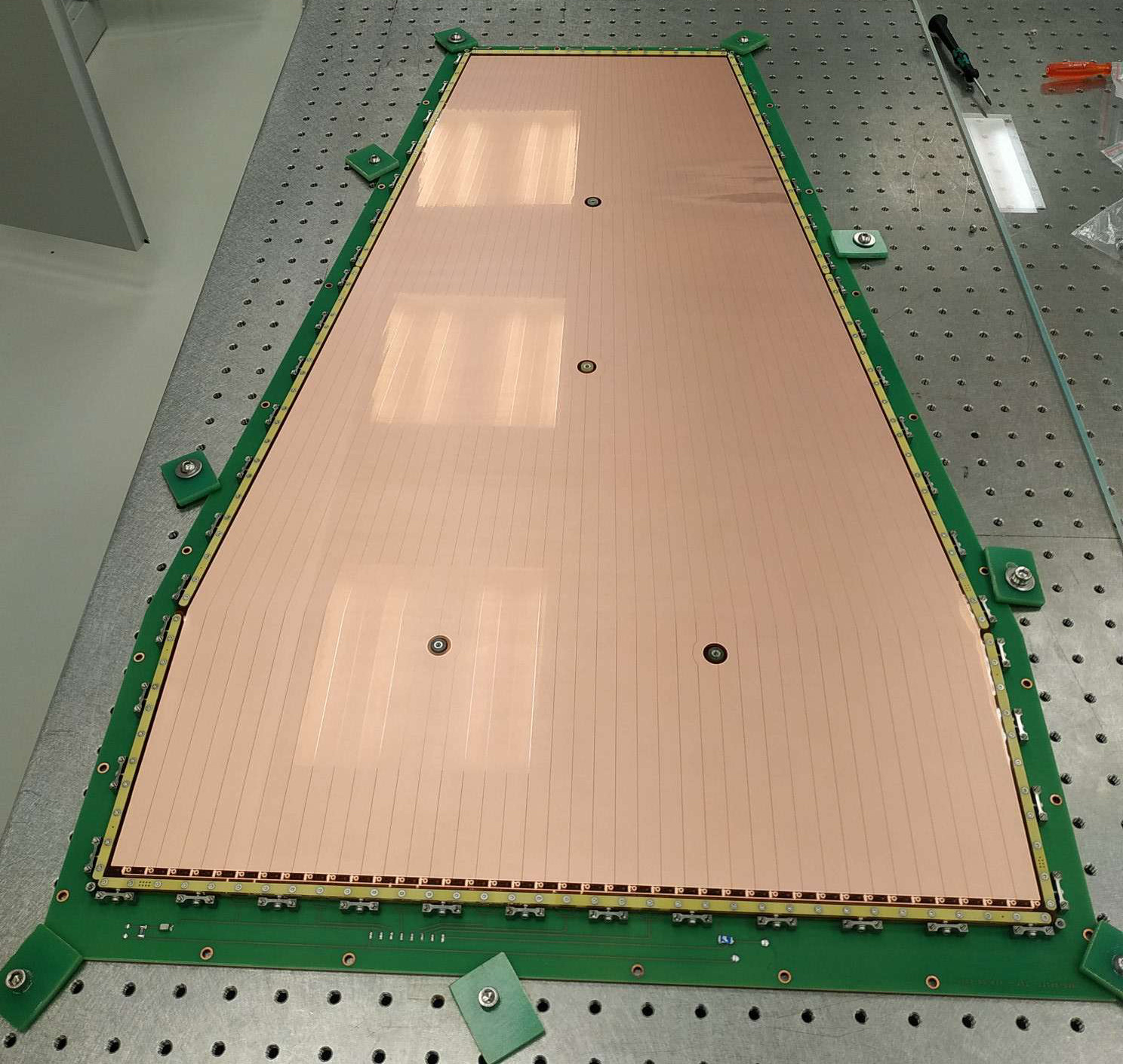}
    \caption{Picture of the first GEM foil produced with azimuthal segmentation.}
    \label{fig:picture_segmentation}
\end{figure}

The future plans for the ME0 GEM detector optimization involve two separate directions. On one hand, more in-depth rate capability studies are ongoing to compare the effect of different choices of protection resistor and of resistive filters. In parallel, a prototype of ME0 detector with azimuthal segmentation has been assembled (Fig.~\ref{fig:picture_segmentation}) and will be tested with x-rays and charged particles to verify its working principle.

\section{Acknowledgements}
The authors would like to thank A. Cardini from INFN sezione di Cagliari for insightful discussions.

\end{document}